\documentclass[epsf,useAMS,usenatbib,usegraphicx]{mn2e}
\usepackage{xcolor}
\usepackage{epsfig}
\usepackage{aas_macros} 
\usepackage{amsmath}
\usepackage{amssymb}
\usepackage{appendix}
\usepackage{booktabs}
\usepackage{upgreek}
\newcommand{\kepler}{\textit{Kepler}}

\title[Phase-modulated stars]{Finding binaries among \textit{Kepler} pulsating stars from phase modulation of their pulsations}

\author[S. J. Murphy et al.] 
{Simon J. Murphy$^{1,2,\dagger}$, Timothy R. Bedding$^{1,2}$, Hiromoto Shibahashi$^3$, \and Donald W. Kurtz$^4$ and Hans Kjeldsen$^{2}$\\
$^1$Sydney Institute for Astronomy (SIfA), School of Physics, University of Sydney, Australia\\
$^2$Stellar Astrophysics Centre, Department of Physics and Astronomy, Aarhus University, 8000 Aarhus C, Denmark\\
$^3$Department of Astronomy, The University of Tokyo, Tokyo 113-0033, Japan\\
$^4$Jeremiah Horrocks Institute, University of Central Lancashire, Preston, PR1 2HE\\
\\
$^{\dagger}$email: murphy@physics.usyd.edu.au
}
 
\begin{document}

\maketitle 

\begin{abstract}
We present a method for finding binaries among pulsating stars that were observed by the Kepler Mission. We use entire four-year light curves to accurately measure the frequencies of the strongest pulsation modes, then track the pulsation phases at those frequencies in 10-d segments. This produces a series of time-delay measurements in which binarity is apparent as a periodic modulation whose amplitude gives the projected light travel time across the orbit.  Fourier analysis of this time-delay curve provides the parameters of the orbit, including the period, eccentricity, angle of ascending node and time of periastron passage.  Differentiating the time-delay curve yields the full radial-velocity curve directly from the \textit{Kepler} photometry, without the need for spectroscopy. We show examples with $\delta$\,Scuti stars having large numbers of pulsation modes, including one system in which both components of the binary are pulsating. The method is straightforward to automate, thus radial velocity curves can be derived for hundreds of non-eclipsing binary stars from \textit{Kepler} photometry alone.
\end{abstract}


\section{Introduction}
\label{sec:intro}

The study of binary stars is fundamental to our understanding of stellar structure and evolution. Eclipsing binaries in particular are the major source of the stellar fundamental parameters mass and radius. In close binary systems, deformation of the stars by tidal forces provides observational constraints on our understanding of tidal interaction. Most stars form in multiple systems; probably $\sim$100\:per\:cent do if we include stars with exoplanets.

Now we have a second, complementary way to derive stellar fundamental parameters: asteroseismology (e.g., \citealt{aertsetal2010}). The study of pulsating stars in binary systems is therefore particularly important (e.g.\ \citealt{handleretal2002,becketal2014}), but most efforts so far have been limited to eclipsing systems (e.g. \citealt{southworthetal2011,hambletonetal2013,debosscheretal2013,frandsenetal2013,maceronietal2014,dasilvaetal2014}).

The clear geometrical information available from the binary observations (light curves and radial velocity curves), combined with the constraints on internal structure from asteroseismology, give us novel insights into stars. Stellar masses, radii and ages are better known than ever before, and physical processes such as convective overshoot, internal rotation, and tidal interaction are now observational sciences. 

The {\it Kepler} mission provides an unprecedented source of high-precision stellar light curves with nearly continuous data over a time span of 4\,yr for over 150\,000 stars. Obtaining the required radial velocity curves for this many stars from ground-based observations is currently not possible. In this paper we show how radial velocities for pulsating binary $\delta$\,Scuti stars observed with {\it Kepler}'s long-cadence mode (30-minute sampling) can be derived from photometry alone. Our technique can be automated to discover and characterise hundreds of binary stars in the {\it Kepler} data set, particularly for non-eclipsing systems that other techniques do not find. It should even be capable of finding planet-mass companions to pulsating stars that are beyond the reach of other surveys for exoplanets.


\section{Method and examples}
\label{sec:tdd}

The orbital motion of a star in a binary system leads to a periodic variation in the path length travelled by the light that we observe. Hence the phase of the observed pulsations varies over the orbit (e.g. \citealt{paparoetal1988,silvottietal2007,silvottietal2011,teltingetal2012}). \citet{shibahashi&kurtz2012} showed that the light-time effect in a binary star leads to a frequency multiplet in the Fourier transform of the light curve of a pulsating star and that the members of the multiplet are separated by the orbital frequency. They demonstrated that the amplitudes and phases of the components of the frequency multiplet can be used to derive all of the information traditionally found from radial velocity curves. The analogy to the frequency modulation (FM) technique found in radio transmitters led to such pulsating stars in binaries being termed `FM stars'. 

The FM method of \citet{shibahashi&kurtz2012} is suitable for the data covering an observational time span that is long compared with the binary orbital period. For wider binaries, however, the FM method is not necessarily appropriate, because the frequency splitting due to the binary motion becomes comparable to the frequency resolution. In this paper, we develop a complementary method, by which we extract explicitly the phase modulation (PM) in the time domain of intrinsic pulsation frequencies of the star, which is caused by binary orbital motion. We exploit this to derive the orbital radial velocities and the other orbital elements from the light curve alone. An advantage of this PM method, compared to the FM method, is that it provides the variation in the light travel time at short intervals. This is particularly useful in visualising the variation caused by the binary motion, and its time derivative provides the radial velocity. Furthermore, the signals from different pulsation modes in multi-periodic stars can be combined. All of this is straightforwardly automated. Unlike the traditional $O-C$ (observed minus calculated) method, PM uses all of the data -- not just the pulsation maxima -- and is particularly suited to multi-mode pulsators.


\subsection{Phase variations and time delays}
\label{ssec:phase_variations}
\label{ssec:tdd}

As a pulsating star moves in its binary orbit, the path length of the light between us and the star varies, leading to the periodic variation in the arrival time of the signal from the star to us. We detected this as a phase variation.

We use entire \kepler\ long-cadence (30-min sampling) data sets, which may be up to 1400\,d in length, for identifying peaks arising from each pulsation mode. For binary stars these peaks are multiplets that are used by the FM method. In our analysis, we measured the frequency of the central component of the multiplet for each peak in the Fourier transform. We then divided the light curve into $n$ short segments and measured the phase at this frequency in each segment. This produced a set of phases as a function of time, which could be repeated for each pulsation mode. Choice of the segment size is a trade-off between time resolution and frequency resolution. We found that 10-d segments gave sufficient frequency resolution to measure phases of well-resolved pulsation modes, while still giving enough time resolution to monitor the time-delays, and ultimately the radial velocities. Shorter segments gave much higher scatter in the radial velocity curve. There is a measured phase, $\phi_{ij}$, for each fixed pulsation frequency $\nu_j$, for each of the time segments $t_i$. If the target is actually a binary system, that series of phases $$\Phi_j (t) = [\phi_{1j}, \phi_{2j}, ~\dots~, \phi_{ij}, ~\dots~, \phi_{nj}]$$varies with the orbital period of the system. We describe the mathematics of the phase variations in Appendix\,\ref{app:mathematics}.

The amplitude of the phase variations depends on the geometry of the orbit, which in turn is dictated by the mass of the perturbing companion, and the frequency of the pulsation (see Eqs.\:\ref{eq:luminosity_variation} and \ref{eq:frequency_dependence} in Appendix\,\ref{app:mathematics}). The latter dependence means that the phase variations of each pulsation mode will have a different amplitude. To remove this effect, we converted the phase variations into light arrival times, which we refer to as time delays. To do this, we first calculated the relative phase shifts by subtracting the mean phase for each mode from each measurement
\begin{equation}
\Delta \phi_{ij} = \phi_{ij} - \overline{\phi_{j}},
\end{equation}
where
\begin{equation}
\overline{\phi_{i}} = \frac{1}{n}\sum_{i=1}^{n}\phi_{ij}.
\end{equation}
Then the time delay with respect to the stationary (single-star) case is simply
\begin{equation}
\label{eq:arrival_time}
\tau_{ij} = \frac{\Delta \phi_{ij}}{2 \uppi \nu_j}.
\end{equation}

In the following, we demonstrate the method with five examples. We show results for each in a figure with four panels: (a) the Fourier transform of the entire light curve, (b) time delays in each segment for the strongest pulsation modes, calculated from the pulsation phases, (c) the Fourier transform of those delays, and (d) the Fourier transform of the weighted average of the time delays.


\subsection{Example 1: a known binary}

\begin{figure*}
\begin{center}
\includegraphics[width=0.92\textwidth]{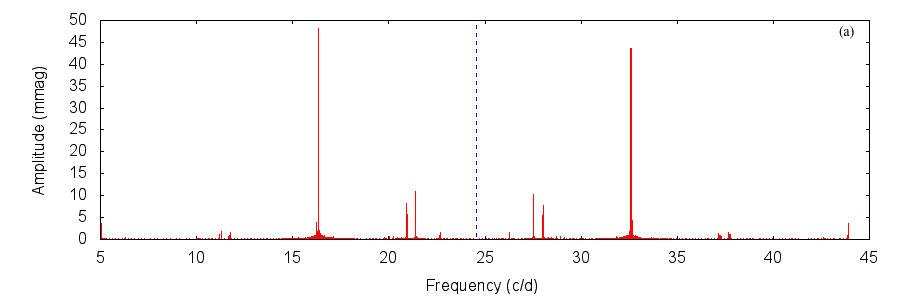}
\includegraphics[width=0.92\textwidth]{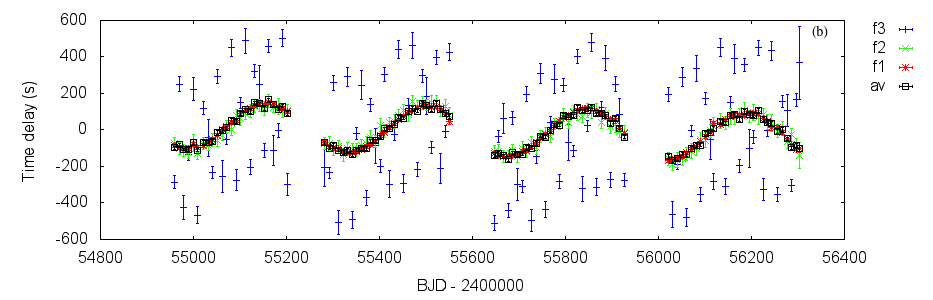}
\includegraphics[width=0.92\textwidth]{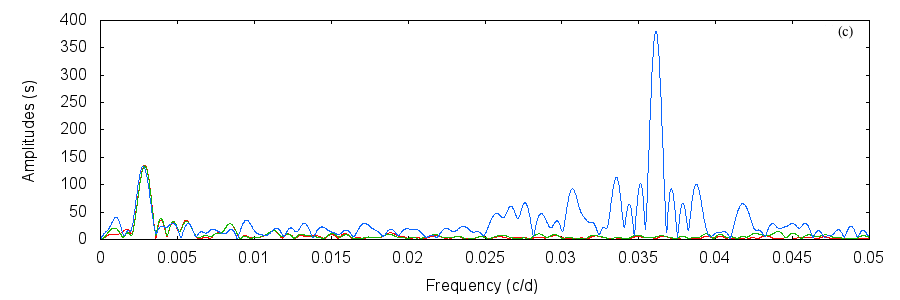}
\includegraphics[width=0.92\textwidth]{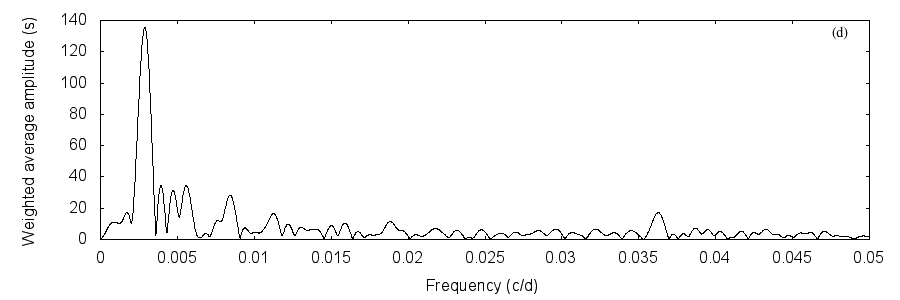}
\caption{Example 1. (a): The Fourier transform of the 1350-d \kepler\ light curve of KIC\,11754974 is shown for both sides of the long-cadence Nyquist frequency (24.47\,d$^{-1}$, blue line), as discussed in \S\,\ref{sssec:frequency_range}. (b): Time delay as a function of time in 10-d segments, for three modes at $f_1 = 16.34$\,d$^{-1}$, $f_2 = 21.40$\,d$^{-1}$, and $f_3 = 20.91$\,d$^{-1}$. The uncertainty-weighted averages of the data are plotted. The orbital variations appear as a sinusoid of amplitude 125\,s and period 344\,d. The gaps arise because the star falls on Module\:3. (c): Fourier transform of the time-delay data in panel (b), with the weighted averages plotted in panel (d). Time delays of all three modes agree well at the orbital frequency at 0.0029\,d$^{-1}$, but an extra peak (and its window pattern) appear for $f_3$ (blue) because of its unresolved frequency. See the text for details.}
\label{fig:11754974}
\end{center}
\end{figure*}

For a binary in a circular orbit, the time delays should follow a sinusoid whose amplitude is the projected light travel time across the orbit and whose period is the orbital period. An example is shown in Fig.\,\ref{fig:11754974}. This star is KIC\,11754974, an SX\,Phe star already demonstrated to be a binary by \citet{murphyetal2013a} based on O$-$C and FM analyses. The star falls on Module\:3 -- a pair of \textit{Kepler}'s CCDs that failed early in the mission, rendering $\sim$4/21 of the field unobservable for 93 consecutive days each year. These outages are of little consequence to our results. Time delays for the three strongest pulsation modes are shown in Fig.\,\ref{fig:11754974}b. The delays for $\nu_1$ and $\nu_2$ agree well and show a clear sinusoidal variation. The delays for $\nu_3$ are noticeably scattered. This results from a pulsation mode close to $\nu_3$ that modifies its phase. In the full 1350-d data set the two frequencies are resolved, but in each 10-d segment they are not, due to a poorer frequency resolution that scales as $1/T$. As such, in the 10-d segments there is beating between these two unresolved frequencies, and the measured phase reflects this. Despite this, the delays for $\nu_3$ still contain information on the orbital period, as evidenced by the Fourier transform of the time-delay data (Fig.\,\ref{fig:11754974}c). Note that lengthening the sampling segments would see the scatter progressively reduced until the segments were long enough that the two frequencies were completely resolved.

All three modes show time delay variations with the same period, amplitude and phase, as seen in their Fourier spectra. The unresolved companion frequency near $\nu_3$ is responsible for the additional blue peak dominating the right of Fig.\,\ref{fig:11754974}c. The separation of the two pulsation frequencies is equal to the frequency of that blue peak: 0.036\,d$^{-1}$. The mathematics of that relationship is described in Appendix\,\ref{app:close_frequency}.

We also calculated the weighted average time delay,
\begin{equation}
\label{eq:weighted_mean}
\langle \tau_i \rangle ~=~ \frac{\sum_{j=1}^{m} \Big[ \frac{\tau_{ij}}{\sigma_{ij}^2}\Big]}{\sum_{j=1}^{m} \Big[ \frac{1}{\sigma_{ij}^2} \Big]},
\end{equation}
weighed by the phase uncertainties. This is shown as black squares in Fig.\,\ref{fig:11754974}b and its Fourier transform is shown in Fig.\,\ref{fig:11754974}d. The derived period of $344.2\pm1.5$\,d is in agreement with the orbital period of $343.3\pm0.3$\,d found by \citet{murphyetal2013a}. The latter period is more precisely determined because those authors also made use of $\sim$1600\,d of data from the Wide Angle Search for Planets (WASP).

\subsubsection{Choice of frequency range for the pulsation spectrum}
\label{sssec:frequency_range}

We calculated the pulsation spectrum in Fig.\,\ref{fig:11754974}a on both sides of the \kepler\ long-cadence Nyquist frequency (24.47\,d$^{-1}$). The real peaks can be distinguished from aliases because the latter are split into multiplets \citep{murphyetal2012b} whose central peaks have lower amplitudes than real peaks. We use the amplitudes to determine which peaks are real. Hence in Fig.\,\ref{fig:11754974}a, we can see that the star's real pulsations lie below the Nyquist frequency.

\begin{figure*}
\begin{center}
\includegraphics[width=0.90\textwidth]{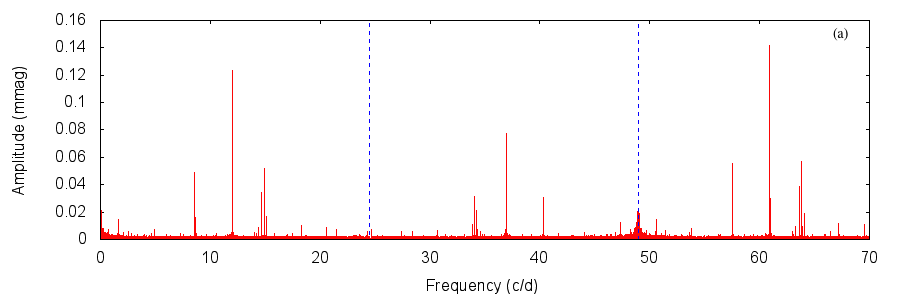}
\includegraphics[width=0.90\textwidth]{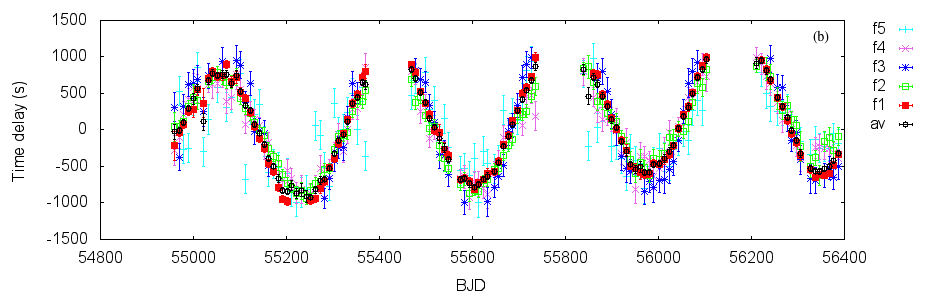}
\includegraphics[width=0.90\textwidth]{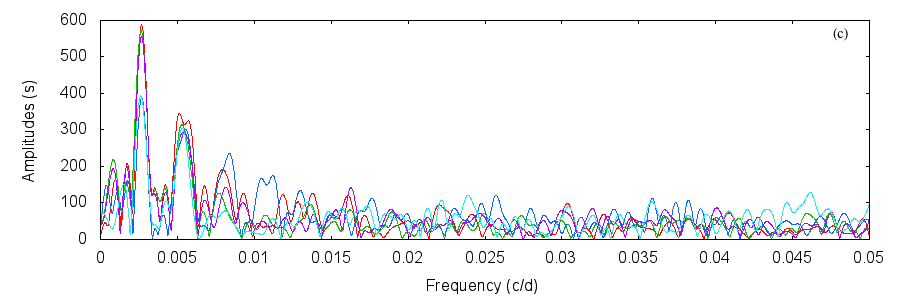}
\includegraphics[width=0.90\textwidth]{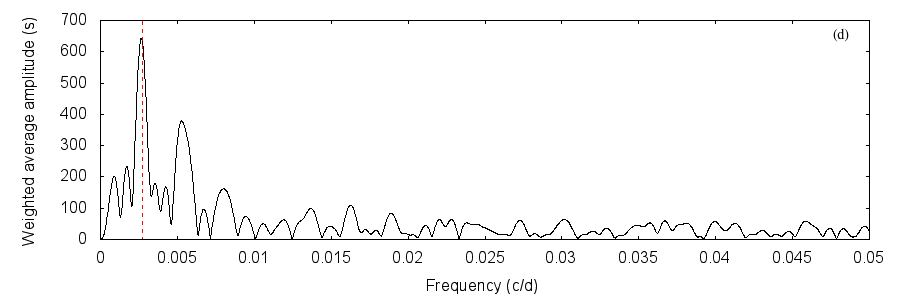}
\caption{An illustration of the spurious signal seen when the phase of a Nyquist alias is tracked instead of a real frequency. (a): The Fourier transform for the non-binary star KIC\,5459908, calculated up to 70\,d$^{-1}$ -- almost three times the Nyquist frequency. Multiples of the Nyquist frequency are indicated by dashed blue lines. Inspection of the peak amplitudes shows the highest peaks occur in the highest-frequency set of peaks. These are therefore the real peaks. We have confirmed this with the Super-Nyquist Asteroseismology method \citep{murphyetal2012b}. However, time delays (b) have been calculated for the lowest-frequency set of peaks, at $f_1 \dots f_5 =$ 12.0, 14.91, 8.6, 14.67 and 15.11\,d$^{-1}$. The time delays for the aliases clearly vary at the \kepler\ orbital frequency. Gaps in the data arise because the star lies on Module\:3. (c): Fourier transform of the time-delay data. (d): Fourier transform of the weighted-average time-delay data, with a dashed red line indicating the orbital frequency of the \kepler\ satellite around the Sun (0.0027\,d$^{-1}$).}
\label{fig:kepler}
\end{center}
\end{figure*}

Tracking a Nyquist alias, rather than the real peak, causes the phases to be modulated at the \kepler\ orbital period (372.5\,d). This is easily detected in our analysis, and is illustrated in Fig.\,\ref{fig:kepler}, for a star that is not binary.

\subsection{Example 2: a long-period binary}
\label{ssec:sensitivity}

The shortest period binary we can detect is set by the length of the sampling segments. In our examples we use 10-d sampling, which limits the sensitivity our application of the method to $P_{\rm orb} > 20$\,d, but shorter segments may be used. Very short-period binaries can be detectable by other means, such as the Fourier series that appear in Fourier transforms of light curves of ellipsoidal variables, and the fact that short-period binaries are often eclipsing. Hundreds of such binaries are already known in the \kepler\ data \citep{slawsonetal2011,gaulmeetal2013}.

\begin{figure*}
\begin{center}
\includegraphics[width=0.92\textwidth]{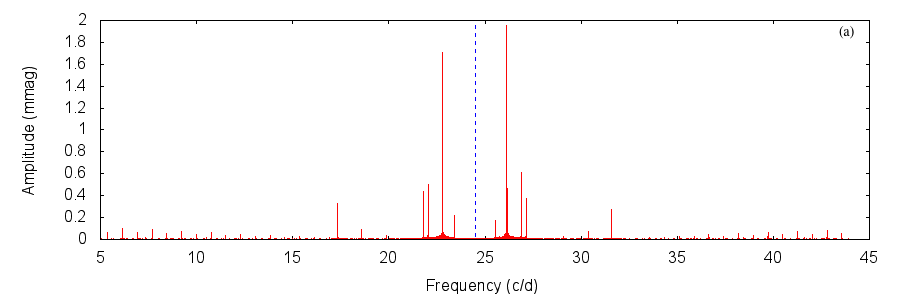}
\includegraphics[width=0.92\textwidth]{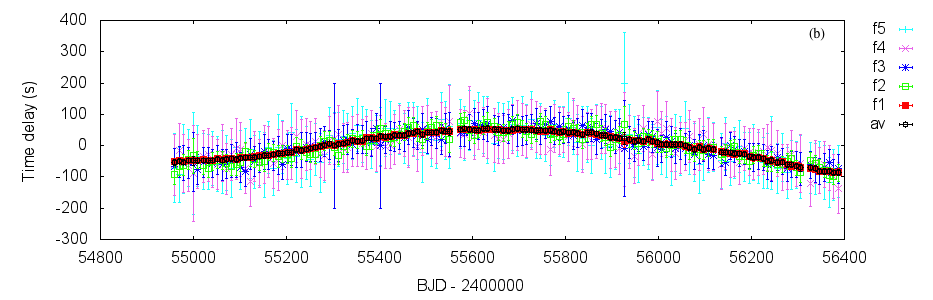}
\includegraphics[width=0.92\textwidth]{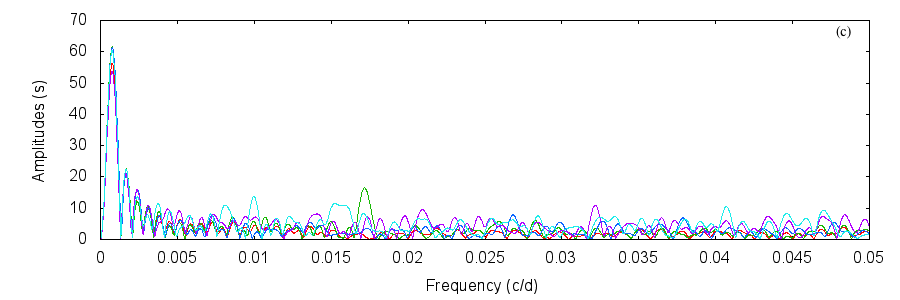}
\includegraphics[width=0.92\textwidth]{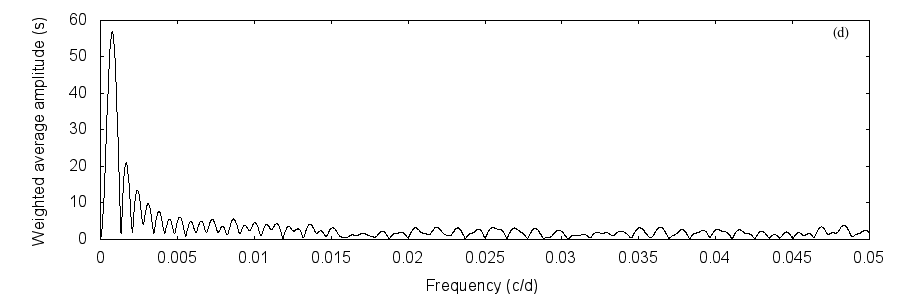}
\caption{Example 2. (a): Fourier transform of the light curve of KIC\,7618364. (b): time delays for the nine highest peaks in panel (a). Their frequencies, in descending amplitude order, are 26.14, 26.9, 21.8, 17.35 and 23.39\,d$^{-1}$, respectively. (c) the Fourier transform of the time-delay data. (d): the Fourier transform of the weighted-average time-delay data.}
\label{fig:7618364}
\end{center}
\end{figure*}

The longest measurable period is set by the length of the data set. It is this requirement -- long, continuous data sets -- that has delayed developments like FM and PM until the present era of space-based photometry, especially that of \textit{Kepler}. Fig.\:\ref{fig:7618364} shows KIC\,7618364, a $\delta$\,Sct star that appears to be in a long-period binary. We find $P_{\rm orb} = 1479\pm17$\,d, which is slightly longer than the \kepler\ data set (1437\,d). We extracted the five highest peaks from its pulsation spectrum (Fig.\,\ref{fig:7618364}a). Note that KIC\,7618364 pulsates at high frequencies, some of which are above the Nyquist frequency of \kepler\ long-cadence data. We see all modes vary with the same periodicity, and the orbital frequency is clearly seen in Figs\,\ref{fig:7618364}c and d.

\subsection{Example 3: a low-amplitude pulsator}
\label{sec:low-amplitude}

It would be useful to be able to apply our method to stars with low pulsation amplitudes. Only 56\:per\:cent of stars in the $\delta$\,Sct instability strip pulsate at amplitudes above 50\,$\upmu$mag \citep{murphy2013}. Among those, stars with peak amplitudes above 0.5\,mmag are outnumbered by those below 0.5\,mmag by almost a factor of three. To maximise the sensitivity of binary searches, our method therefore needs to be capable of detecting binaries among the low-amplitude pulsators.

\begin{figure*}
\begin{center}
\includegraphics[width=0.92\textwidth]{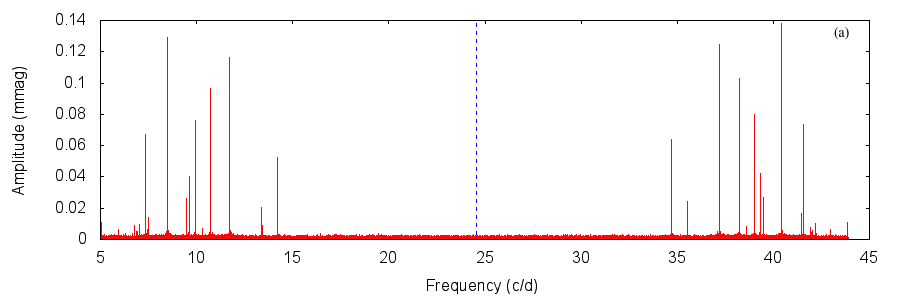}
\includegraphics[width=0.92\textwidth]{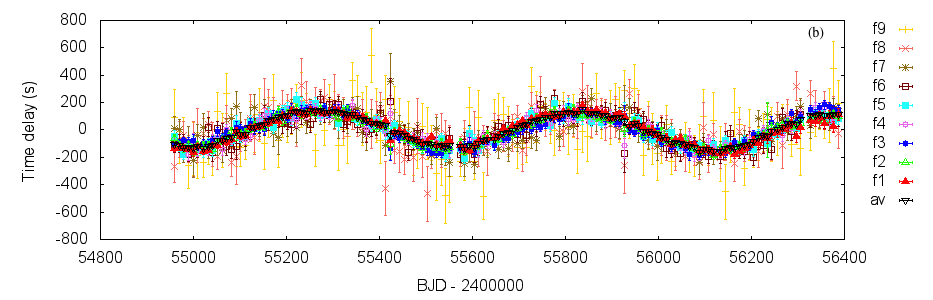}
\includegraphics[width=0.92\textwidth]{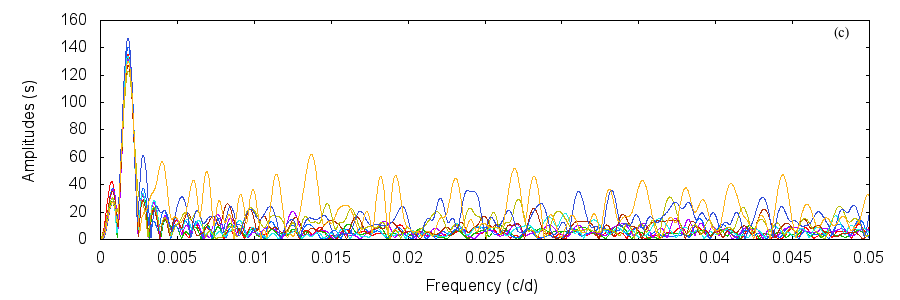}
\includegraphics[width=0.92\textwidth]{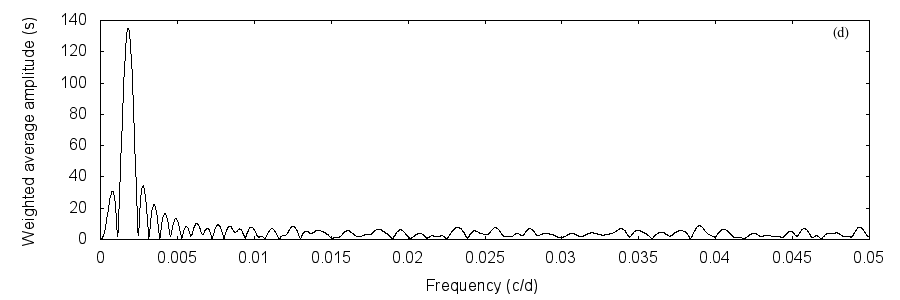}
\caption{Example 3. (a): Fourier transform of the light curve of the low amplitude $\delta$\,Sct star KIC\,11771670. (b): light arrival time delays for the highest nine peaks from panel (a). Their frequencies, in descending amplitude order, are 40.43, 37.21, 38.23, 39.01, 41.58, 34.71, 39.33, 39.49 and 35.54\,d$^{-1}$, respectively. (c): Fourier transforms of the time delays for those nine peaks. (d): the Fourier transform of the weighted-average time-delay data, giving $P_{\rm orb} = 565.65\pm1.83$\,d.}
\label{fig:11771670}
\end{center}
\end{figure*}

To test this, Fig.\,\ref{fig:11771670} shows the $\delta$\,Sct star KIC\,11771670, whose pulsation amplitudes range from 0.12\,mmag downwards. Its pulsation spectrum is displayed in Fig.\,\ref{fig:11771670}a. From the argument presented in \S\,\ref{sssec:frequency_range}, we conclude that the real pulsation frequencies of KIC\,11771670 lie above the long-cadence Nyquist frequency. We plot their time delays in Fig.\,\ref{fig:11771670}b. Each mode shows a clear variation at the orbital period (565\,d) and the orbital information can be extracted even for the weakest mode monitored (0.02\,mmag). From this we conclude that binary companions are detectable for even the lowest-amplitude $\delta$\,Sct stars.

\subsection{Example 4: an eccentric binary}
\label{ssec:eccentricity}

\begin{figure*}
\begin{center}
\includegraphics[width=0.92\textwidth]{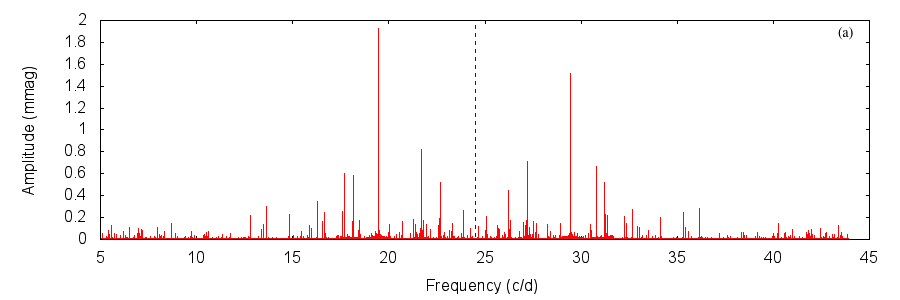}
\includegraphics[width=0.92\textwidth]{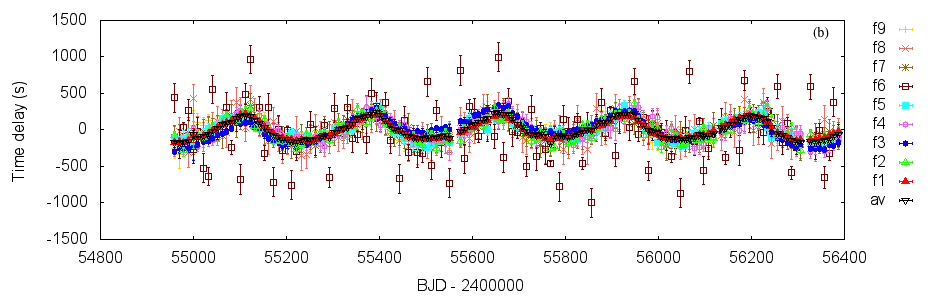}
\includegraphics[width=0.92\textwidth]{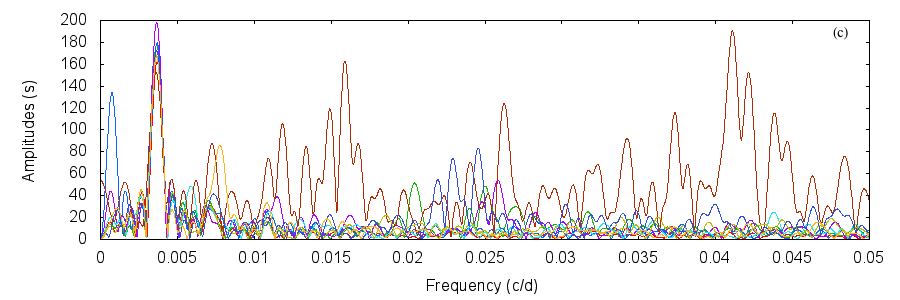}
\includegraphics[width=0.92\textwidth]{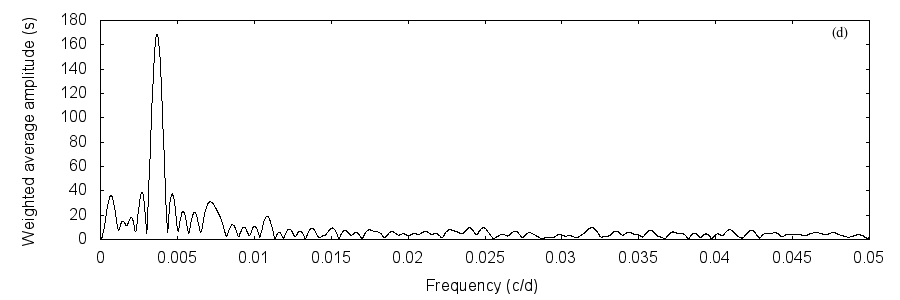}
\caption{Example 4. (a): The Fourier transform of the light curve of KIC\,9651065. (b): Time delays for the nine-highest peaks. Their frequencies, in descending amplitude order, are 19.48, 21.71, 30.8, 17.7, 22.69, 24.46, 16.27, 13.62 and 36.15\,d$^{-1}$, respectively. (c): Fourier transforms of the time delays for those nine peaks. (d): Fourier transform of the weighted-average time-delay data. Panel (d) shows the main peak and two harmonics. The amplitudes of those harmonics, $A_1=165.6$, $A_2 = 38.4$ and $A_3 = 10.2$\,s ($\pm2$\,s), are obtained from a least-squares fit at $\nu_{\rm orb}$, $2\nu_{\rm orb}$ and $3\nu_{\rm orb}$.}
\label{fig:9651065}
\end{center}
\end{figure*}

An eccentric binary will produce a non-sinusoidal time-delay curve and we expect harmonics of the orbital frequency in its Fourier transform. Fig.\,\ref{fig:9651065} shows KIC\,9651065, which was discovered to be highly eccentric by \textbf{Shibahashi et al. (in prep.)}. The time delays (Fig.\,\ref{fig:9651065}b) are clearly not represented by a single sinusoid, but can be described by a Fourier series. The Fourier transforms (Figs\,\ref{fig:9651065}c and d) show the main peak and the first two harmonics. The eccentricity (and $\varpi$, the angle between the ascending node and the periapsis) can be determined from the amplitude ratios of the harmonics, just as they are determined from the amplitude ratios of consecutive sidelobes in the frequency domain (cf. equation 46 of \citealt{shibahashi&kurtz2012}). Particularly, in the case of $e \ll 1$,
\begin{equation}
\label{eq:eccentricity}
e \simeq 2A_2 / A_1,
\end{equation}
where $A_1$ and $A_2$ are the amplitudes of the first and second harmonics, respectively, that is, the amplitudes of peaks at $\nu_{\rm orb}$ and $2\nu_{\rm orb}$. In the more general case, we can determine $e$ and $\varpi$ from the ratios $A_2/A_1$ and $A_3/A_2$, which we defer to Appendix\:\ref{app:eccentricity} along with the derivation of Eq.\,\ref{eq:eccentricity}.

\subsection{The radial velocity curve}
\label{ssec:rv}

Time delays are caused by the modulation of the length of the light path. Hence they are described by the integral
\begin{equation}
\tau (t) = -\frac{1}{c}\int_0^t v_{\rm rad,1}(t')\,dt'.
\label{eq:13}
\end{equation}
The time derivative of the time delays gives the radial velocity as a function of time:
\begin{equation}
v_{\rm rad,1}(t) = -c{{d \tau}\over{dt}}.
\label{eq:rv}
\end{equation}

\begin{figure*}
\begin{center}
\includegraphics[width=0.92\textwidth]{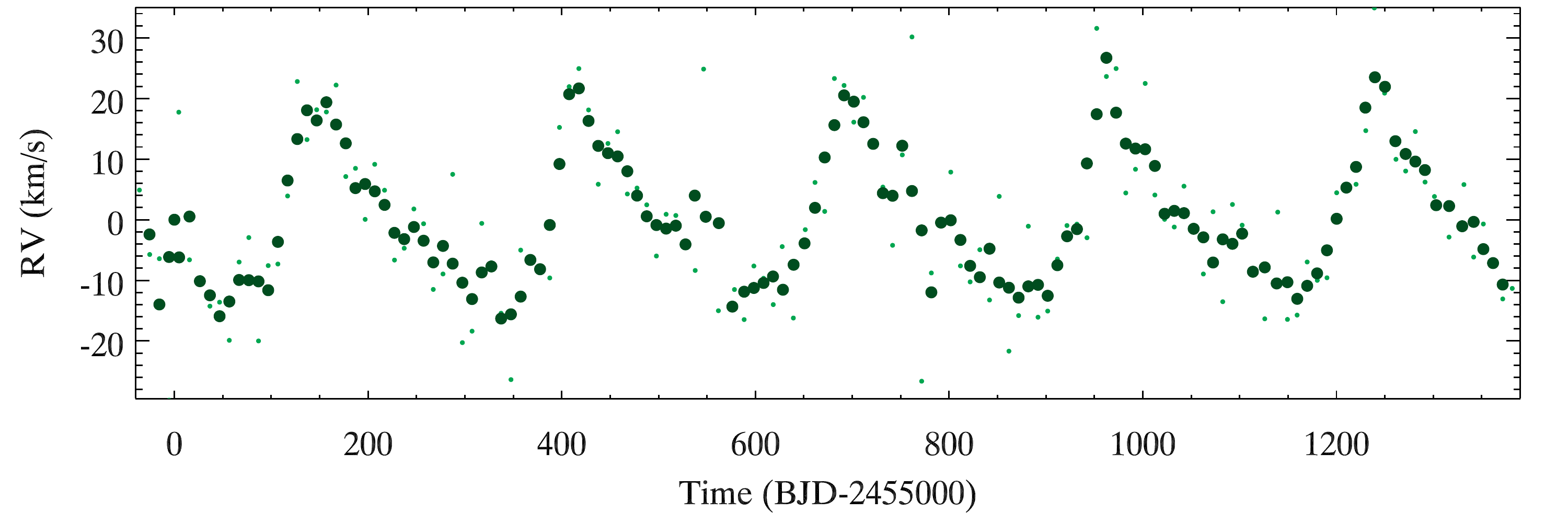}
\includegraphics[width=0.92\textwidth]{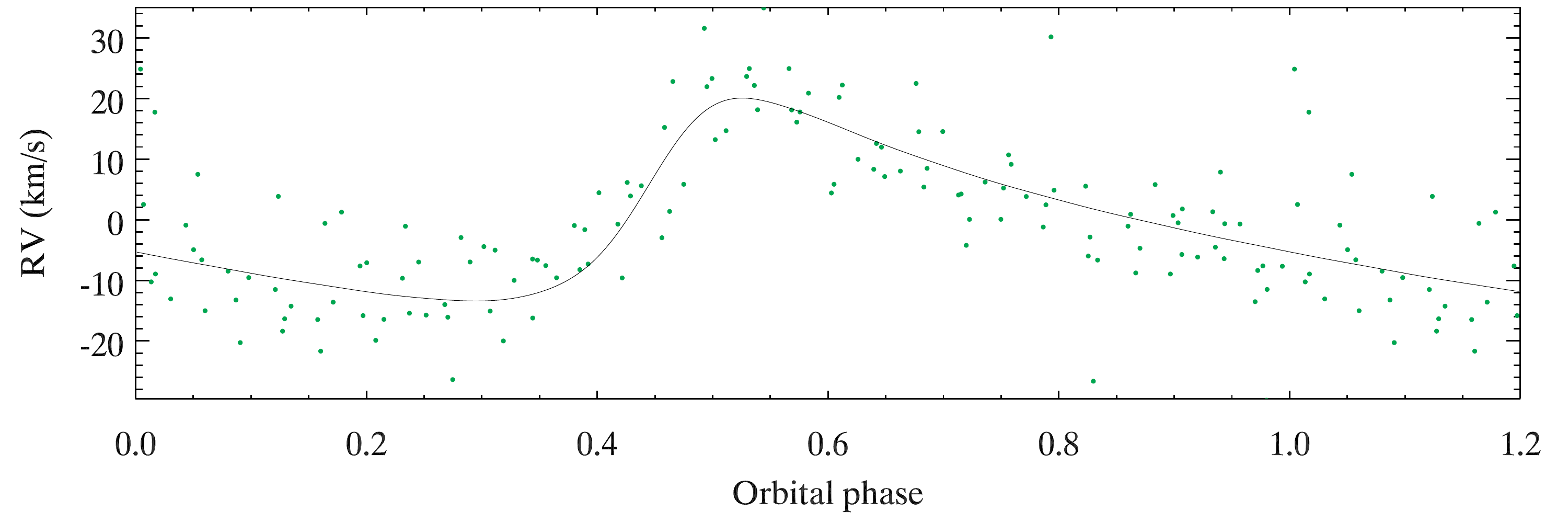}
\caption{Radial velocity curve for KIC\,9651065 as a function of time (top) and phased on $P_{\rm orb} = 272.7$\,d (bottom). The zero-point of the phased radial velocities was chosen to coincide with the start of the \kepler\ data set. The radial velocities of each 10-d light-curve segment from pair-wise differences are shown as small circles; large circles represent smoothing of those values over three points. The black line is the RV curve calculated from the differentiated analytical function.}
\label{fig:rv}
\end{center}
\end{figure*}

One way to differentiate the data is numerically. Pair-wise differentials in time delays over each light curve segment give the average radial velocity in those segments. To reduce the scatter, we smoothed over three consecutive time delays using a moving boxcar function. The pair-wise differentials and the smoothed values are displayed in Fig.\,\ref{fig:rv}, for the eccentric star discussed in the previous example (KIC\,9651065).

A second method uses an analytical approach to produce a smooth RV curve (black line, Fig.\,\ref{fig:rv}). For a circular orbit, the time delay is described by the sinusoid
\begin{equation}
\tau (t) = A \sin{}(2 \uppi \nu_{\rm orb} t + \phi),
\label{eq:porb}
\end{equation}
where $A$ is the amplitude (the maximum time delay), $\nu_{\rm orb}$ is the orbital frequency (reciprocal of the orbital period, $P_{\rm orb}$), and $\phi$ is the phase with respect to the chosen zero-point in time. Eccentric orbits are described by a Fourier series, that is, a superposition of $N$ harmonics of the orbital frequency
\begin{equation}
\tau (t) = \sum_{k=1}^N A_k \sin{} (2 \uppi k \nu_{\rm orb}t + \phi_k).
\label{eq:fourier_terms}
\end{equation}
The orbital harmonics in the time delays are fitted with least-squares to find their amplitudes: $A_1=165.6$, $A_2 = 38.4$ and $A_3 = 10.2$\,s, each with a formal uncertainty of $\pm$2.0\,s. The eccentricity ($e = 0.468 \pm 0.025$) is calculated with the help of Bessell functions (cf. Appendix\:\ref{app:eccentricity}) using ratios of those amplitudes. The full orbital solution, including details on the RV curve, is discussed in Appendix\:\ref{app:calculation}.

RV curves derived with the PM method could be used as input for codes that model eccentric binaries, such as {\small PHOEBE} \citep{prsa&zwitter2005}. Given that such codes aim to infer the geometry of the orbit, modelling the time delays themselves might be preferred over the RV curve, since the former give the binary geometry directly and more precisely.


\subsection{Orbital parameters}

In this section, we explain how information on the binary system can be extracted from the data. Obtaining the eccentricity and orbital period has already been discussed (Eqs\,\ref{eq:eccentricity} and \ref{eq:porb}). In the circular case the projected semi-major axis of the pulsator's orbit around the centre of mass is simply the maximum time delay, $A$, times the speed of light:
\begin{equation}
\label{eq:a1sini}
a_1 \sin i = Ac,
\end{equation}
but the eccentric case requires a more detailed treatment.

To first order, the radial velocity curve of the pulsating star is described by Eqs\:\ref{eq:rv} and \ref{eq:fourier_terms}. In the general case, the radial velocity $v_{\rm rad,1}$ is expressed as
\begin{equation}
v_{\rm rad,1}
=
-{{2\uppi\nu_{\rm orb} a_1\sin i}\over{\sqrt{1-e^2}}}
\left[
\cos(f+\varpi) + e\cos\varpi
\right],
\end{equation}
where $a_1$ denotes the semi-major axis, $e$ is the eccentricity, $f$ is the true anomaly, $\varpi$ denotes the angle between the ascending node and the periapsis, and $i$ denotes the inclination angle of the orbital axis with respect to the line of sight. The angle $f+\varpi$ defines the angle between the ascending node and the star at the time when $v_{\rm rad,1}$ is measured. This value of $v_{\rm rad,1}$ is the same quantity that would be measured by spectroscopy.

The minimum and the maximum values of the radial velocity are then
\begin{eqnarray}
v_{\rm rad,1,min} \equiv 
-{{2\uppi\nu_{\rm orb} a_1\sin i}\over{\sqrt{1-e^2}}}
\left(
1 + e\cos\varpi
\right)
\end{eqnarray}
and 
\begin{eqnarray}
v_{\rm rad,1,max} \equiv 
-{{2\uppi\nu_{\rm orb} a_1\sin i}\over{\sqrt{1-e^2}}}
\left(
-1 + e\cos\varpi
\right),
\end{eqnarray}
respectively,
and they are available from the radial velocity derived from the time delays.
Then, the projected semi-major axis of the pulsator's orbit around the barycentre is deduced:
\begin{equation}
a_1\sin i = {{1}\over{4\uppi}} {{1}\over{\nu_{\rm orb}}} \sqrt{1-e^2} 
\left( v_{\rm rad,1,max} - v_{\rm rad,1,min} \right),
\end{equation}
with the help of the orbital frequency and the eccentricity already derived from the amplitude ratio of the harmonics in the non-sinusoidal time-delay curve (Eq.\,\ref{eq:eccentricity}).
The value of $a_1\sin i$ obtained in this way is used to calculate the mass function:
\begin{align}
\label{eq:mass_fn}
f(m_1, m_2, \sin i)& = \frac{(m_2 \sin i)^3}{(m_1 + m_2)^2}\\
& = \frac{(2 \uppi)^2}{P_{\rm orb}^2 G} ~ (a_1 \sin i)^3,
\end{align}
where $G$ is the gravitational constant. Thus with a suitable assumption of the primary mass (or, even better, its mass as determined through asteroseismology), the mass of the perturbing body, $m_2$, can also be found.

The angle between the ascending node and the periapsis, $\varpi$, is
\begin{equation}
\cos\varpi = 
-\frac{1}{e}
\left[
{{v_{\rm rad,1,max} + v_{\rm rad,1,min}}\over{v_{\rm rad,1,max} - v_{\rm rad,1,min}}}
\right].
\label{eq:angle}
\end{equation}
This relation can be used to confirm those values deduced from the amplitude ratios of the harmonics of the orbital period in the time delays.

When the star passes the periastron, the true anomaly is zero ($f=0$), hence
\begin{equation}
v_{\rm rad, 1, periastron} = {{1+e}\over{2 e}} 
\left( v_{\rm rad,1,max} + v_{\rm rad,1,min} \right).
\end{equation}
From this value and the radial velocity curve as a function of time, we can derive the time of periastron passage of the star.


\subsection{Example 5: a binary with two pulsating components}

\begin{figure*}
\begin{center}
\includegraphics[width=0.92\textwidth]{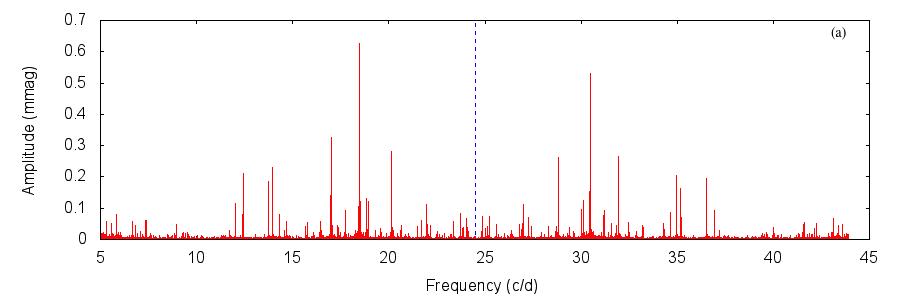}
\includegraphics[width=0.92\textwidth]{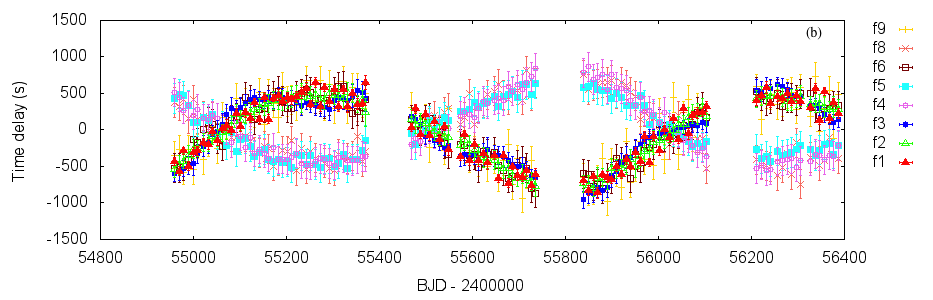}
\includegraphics[width=0.92\textwidth]{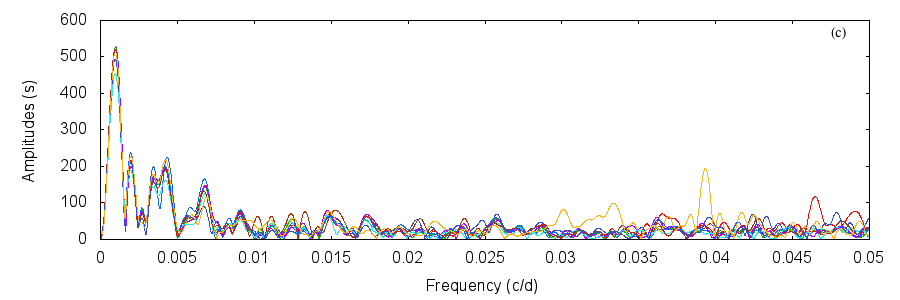}
\caption{Example 5. (a): Fourier transform of the light curve of the $\delta$\,Sct star KIC\,4471379. (b): time delays for the highest nine peaks from panel (a), whose variations as two sinusoids of opposite phase show they belong to two separate stars. (c): Fourier transforms of the time delays for those nine peaks, giving $P_{\rm orb} = 960\pm12$\,d. Unlike in other figures, we do not show the weighted average for obvious reasons. The frequencies of $f_1 \dots f_9$ are 18.45, 16.99, 20.13, 13.96, 12.41, 13.74, 18.83, 12.01 and 21.95\,d$^{-1}$, respectively.}
\label{fig:4471379}
\end{center}
\end{figure*}

Fig.\,\ref{fig:4471379} shows a binary with $P_{\rm orb} = 961\pm12$\,d, KIC\,4471379, consisting of two pulsating stars. We call this a PB2 system, by analogy with the double-lined spectroscopic binary (``SB2'') systems of spectroscopy.

The Fourier transform of the light curve (Fig.\,\ref{fig:4471379}a) shows a dense pulsation spectrum. Fig.\,\ref{fig:4471379}b shows this is the result of the combined light variations of two stars. We see two sets of modes, one set for each star, each having the same period and amplitude. However, the two sets are in anti-phase with each other. The technique has separated the pulsation spectra of two pulsating stars in a binary system very clearly. More detailed analysis of this system will be the subject of a future paper.

\section{Conclusions and Future Work}
\label{sec:conclusions}

The long data sets with high duty cycle provided by \kepler\ have allowed the detection of binary companions to pulsating stars through phase modulation (PM) of the pulsation modes. As with the FM method \citep{shibahashi&kurtz2012}, binary information can be extracted from the \kepler\ light curve without the need for spectroscopy. Furthermore, the PM method is easily automated, offers a clear visualisation of the binary geometry, and straightforwardly provides the instantaneous radial velocity curve. The choice of FM or PM analysis is to be made on a case-by-case basis, and the methods are complimentary. While FM performs more satisfactorily for short-period binaries, where PM sampling segments must be short and thus suffer poor frequency resolution, PM is preferable for long-period binaries and produces a more detectable signal than FM for low-frequency pulsations.

We have shown that the PM method allows companions to be found in orbits at least as long as the 1400-d \kepler\ data sets. Moreover, companions can be found around very low amplitude ($\sim$$20$\,$\upmu$mag) pulsators, regardless of orbital period. Extrapolating this finding, we infer that planetary companions could be discovered around mmag amplitude pulsators. We will be applying this analysis to the full set of $\delta$\,Sct stars in the \kepler\ archive.

\section*{Acknowledgements}
This research was supported by the Australian Research Council. Funding for the Stellar Astrophysics Centre is provided by The Danish National Research Foundation (grant agreement no.: DNRF106). The research is supported by the ASTERISK project (ASTERoseismic Investigations with SONG and Kepler) funded by the European Research Council (grant agreement no.: 267864). We are grateful to the entire \kepler\ team for such exquisite data. We would like to thank the anonymous referee for helping us make some clarifications.

\bibliography{arxiv_pm}

\appendix


\section{Mathematical derivations}

\subsection{Mathematics of the phase variations}
\label{app:mathematics}

We consider a star pulsating with frequency $\nu_0$ in a binary with circular orbital motion of frequency $\nu_{\rm orb}$. The observed luminosity as a function of time $t$ has the form
\begin{eqnarray}
L(t) &=& A \exp i \left\{2 \uppi \nu_0 
\left[ t- {{1}\over{c}} \int_0^t  v_{\rm rad,1}(t') \,dt' \right]+ \phi \right\} \label{eq:luminosity_variation} \\
&=&
A\exp i\left[(2 \uppi \nu_0 t+\phi) - \alpha\sin{}(2 \uppi \nu_{\rm orb} t\right)],
\label{eq:frequency_dependence}
\end{eqnarray} 
where $\alpha$ is the amplitude of the phase modulation. \citet{shibahashi&kurtz2012} defined the instantaneous frequency from the time derivative of $L(t)$,
\begin{equation}
\nu_{\rm obs} (t)= 2 \uppi \nu_0 \left[ 1- {{v_{\rm rad,1(t)}}\over{c}} \right].
\end{equation}
Similarly, by considering the frequency $\nu_0$ fixed, we can define the instantaneous phase as a slowly varying function of time:
\begin{equation}
\Phi (t) = -\alpha\sin{}(2 \uppi \nu_{\rm orb} t) + \phi.
\label{eq:3}
\end{equation}
Even though the amplitude of the Doppler frequency shift, $\alpha\nu_{\rm orb}/\nu_0$ is tiny, $\alpha$ is usually much bigger.
\citet{shibahashi&kurtz2012} carried out the Fourier transform by implicitly assuming the time span is infinitely long, and then 
\begin{equation}
L(t) 
=
A \sum_{n=-\infty}^{\infty} J_n(\alpha) \exp i2\uppi\left[(\nu_0 - n\nu_{\rm orb}) t + \phi\right].
\end{equation}

In practice, the time span is finite. For the method presented in \S\,\ref{sec:tdd}, we divide the observational time span into equal segments, whose lengths are much longer than the pulsation period $1/\nu_0$ but much shorter than the orbital period $1/\nu_{\rm orb}$. Then a least-squares fit at frequency $\nu_0$ is carried out in each segment. Since the instantaneous phase $\Phi$ is almost constant in each segment, the result is
\begin{align}
\int_{t_n}^{t_{n+1}} &L(t) \exp (-i2\uppi\nu t)\,dt \nonumber\\
& \simeq e^{ i\Phi (t_{n+1/2})} \int_{t_n}^{t_{n+1}} A \exp i 2 \uppi( \nu_0 - \nu_{\rm orb})t \,dt\\
& \simeq e^{ i\Phi (t_{n+1/2})} A {{\Delta t}\over{2}}{\rm sinc}{{2\uppi\Delta\nu\Delta t}\over{2}}
e^{i2\uppi\Delta\nu t_{n+1/2}},
\label{eq:5}
\end{align}
and we get the peak frequency at $\nu=\nu_0$, the amplitude $A$, and the instantaneous phase $\Phi(t_{n+1/2})$. 
Note that the frequency thus obtained is common in all the segments, while the phase $\Phi$ is not, that is, $\Phi$ is a slowly varying and discrete function of time.
Since $\Phi$ is of the form of Eq. (\ref{eq:3}), the binary information, $\alpha$ and $\nu_{\rm orb}$, can be extracted as in the FM method.


\subsection{Closely spaced modes}
\label{app:close_frequency}
Let us assume two unresolved modes are present at $\nu_0$ and $\nu_0 + \delta\nu$, with the same amplitude, for simplicity:
\begin{align}
L(t) =& Ae^{i2\uppi\nu_0\left(t-c^{-1}\int v_{\rm rad,1}\,dt'\right)}
+
Ae^{i2\uppi(\nu_0+\delta\nu)\left(t-c^{-1}\int v_{\rm rad,1}\,dt'\right)}\\
\simeq&
A\exp{[i2\uppi(\nu_0+\delta\nu/2)t]} \nonumber \\
\times&
\exp\left[-i2\uppi(\nu_0+\delta\nu/2)c^{-1}\int v_{\rm rad,1}\,dt'\right] \nonumber \\
\times&
2\cos{}(\uppi\delta\nu)\left(t-c^{-1}\int v_{\rm rad,1}\,dt'\right).
\end{align}
The first term on the right-hand side gives the frequency $\nu_0+\delta\nu/2$, which is $\nu_3$ in our example in \S\,\ref{ssec:tdd}.
The second term gives the phase modulation due to binarity, which coincides with those of $\nu_1$ and $\nu_2$.
The third term indicates the existence of a peak at $\delta\nu/2$ in the power spectrum of the phase of the $\nu_3$-component, with side lobes separated from the peak by exactly $\nu_{\rm orb}$.


\subsection{Measurement of the orbital elements through PM}
\label{app:eccentricity}

In a general case, the radial velocity $v_{\rm rad,1}$ is expressed as
\begin{equation}
v_{\rm rad,1}
=
-{{2\uppi\nu_{\rm orb} a_1\sin i}\over{\sqrt{1-e^2}}}
\left[
\cos(f+\varpi) + e\cos\varpi
\right],
\end{equation}
where $a_1$ denotes the semi-major axis, $e$ is the eccentricity, $f$ is the true anomaly, $\varpi$ denotes the angle between the ascending node and the periapsis, and $i$ denotes the inclination angle of the orbital axis with respect to the line of sight. The angle $f+\varpi$ defines the angle between the ascending node and the star at the moment.

The trigonometric functions of the true anomaly $f$ are expressed in terms of the mean anomaly $l$:
\begin{equation}
\cos f = -e + {{2(1-e^2)}\over{e}}\sum_{n=1}^\infty J_n(ne)\cos nl,
\end{equation}
\begin{equation}
\sin f = 2\sqrt{1-e^2} \sum_{n=1}^\infty J_n{'}(ne) \sin nl.
\end{equation}
Here $J_n(x)$ denotes the Bessel function of the first kind of integer order $n$ and $J_n{'}(x)$ is its derivative ($dJ_n(x)/dx$). With the help of these expressions and the series of $a_n(e)$, $b_n(e)$ and $\xi_n(e,\varpi)$ introduced by 
\citet[][see their figure 11]{shibahashi&kurtz2012}, the radial velocity is rewritten as
\begin{align}
v_{\rm rad, 1}
=
-{2\uppi\nu_{\rm orb} a_1\sin i}
& \left[
\sum_{n=1}^\infty n a_n(e) \cos nl \cos\varpi \right. \nonumber \\
&\left. - \sum_{n=1}^\infty n b_n(e) \sin nl \sin\varpi
\right] \nonumber \\ \nonumber
\end{align}
\begin{equation}
=
-{2\uppi\nu_{\rm orb} a_1\sin i}
\sum_{n=1}^\infty n \xi_n(e,\varpi) \cos \left[ nl + \vartheta_n (e, \varpi) \right],
\label{eq:18}
\end{equation}
where
\begin{eqnarray}
a_n(e) \equiv {{2\sqrt{1-e^2}}\over{e}} {{1}\over{n}} J_n(ne),
\end{eqnarray}
\begin{eqnarray}
b_n(e) \equiv {{2}\over{n}} J_n{'}(ne),
\end{eqnarray}
\begin{eqnarray}
\label{eq:xi}
\xi_n(e,\varpi) \equiv \sqrt{ \{a_n(e)\}^2 \cos^2\varpi + \{b_n(e)\}^2 \sin^2\varpi },
\end{eqnarray}
and
\begin{eqnarray}
\vartheta_n(e,\varpi)
\equiv
\tan^{-1} \left[ {{b_n(e)}\over{a_n(e)}} \tan\varpi \right] .
\end{eqnarray}

Since the mean anomaly $l$ is proportional to the time after the periapsis passage, $t$,
\begin{equation}
l = 2\uppi\nu_{\rm orb} t,
\end{equation} 
equation (\ref{eq:18}) is written explicitly as a function of $t$:
\begin{align}
v_{\rm rad, 1}(t) 
=&
-{2\uppi\nu_{\rm orb} a_1\sin i}\nonumber \\
&\times \sum_{n=1}^\infty n \xi_n(e,\varpi) \cos \left[ 2\uppi n \nu_{\rm orb} t + \vartheta_n (e, \varpi) \right].
\end{align}
Integrating this, we eventually obtain
\begin{align}
&\tau (t) = -{{1}\over{c}} \int_0^t v_{\rm rad, 1}(t')\,dt'
\nonumber \\
&= {{1}\over{c}} {a_1\sin i} \sum_{n=1}^\infty \xi_n(e,\varpi) 
\left[
\sin (2\uppi n\nu_{\rm orb} t + \vartheta_n) - \sin\vartheta_n 
\right] 
\\
&=
{{1}\over{c}} {a_1\sin i}
\left[
\sum_{n=1}^\infty \xi_n(e,\varpi) 
\sin (2\uppi n\nu_{\rm orb} t + \vartheta_n)
+\tau_0(e,\varpi)
\right]  ,
\label{eq:26}
\end{align}
where
\begin{eqnarray}
\tau_0(e,\varpi) \equiv -\sum_{n=0}^\infty \xi_n(e,\varpi)\sin\vartheta_n(e,\varpi) .
\end{eqnarray}

Equation (\ref{eq:26}) means that the amplitude of the $n$-th order Fourier component of the time delay, $A_n$, is 
\begin{eqnarray}
A_n = {{1}\over{c}} a_1\sin i \xi_n(e, \varpi).
\label{eq:28}
\end{eqnarray}
By measuring PM, we know the values of $\{A_n\}$, for $n=1, 2, 3, \dots$. 
Hence from a set of equation (\ref{eq:28}) for $n=1, 2, 3$, we can deduce the eccentricity $e$, the semi-major axis $a_1$, and the angle between the ascending node and the periapsis, $\varpi$.
Indeed, from the ratio of $A_2/A_1$ and $A_3/A_2$, $e$ and $\varpi$ are deduced:
\begin{eqnarray}
{{A_2}\over{A_1}} = {{\xi_2(e,\varpi)}\over{\xi_1(e,\varpi)}}
\label{eq:29}
\end{eqnarray}
and
\begin{eqnarray}
{{A_3}\over{A_2}} = {{\xi_3(e,\varpi)}\over{\xi_2(e,\varpi)}}.
\label{eq:30}
\end{eqnarray}


\subsection{Measurement of eccentricity}
\label{app:a4}

To the order of $O(e^6)$, the coefficients $a_n(e)$ and $b_n(e)$ are given by
\begin{eqnarray}
a_1(e)
\simeq
\sqrt{1-e^2}
\left(
1-{{1}\over{8}}e^2 +{{1}\over{192}}e^4 -{{1}\over{9216}}e^6
\right)
\end{eqnarray}
and
\begin{eqnarray}
b_1(e)
\simeq
1-{{3}\over{8}}e^2 +{{5}\over{192}}e^4 -{{7}\over{9216}}e^6,
\end{eqnarray}
respectively \citep{shibahashi&kurtz2012}. Also,
\begin{eqnarray}
a_2(e)
\simeq
{{e}\over{2}}\sqrt{1-e^2}
\left(
1-{{1}\over{3}}e^2 +{{1}\over{24}}e^4 -{{1}\over{360}}e^6
\right)
\end{eqnarray}
and
\begin{eqnarray}
b_2(e)
\simeq
{{e}\over{2}}
\left(
1-{{2}\over{3}}e^2 +{{1}\over{8}}e^4 -{{1}\over{90}}e^6
\right).
\end{eqnarray}

So, in the case of $e \ll 1$, to the lowest order of $e$, $a_1(e)\simeq 1$ and $b_1(e)\simeq 1$, then $\xi_1(e, \varpi)\simeq 1$.
Similarly,
$a_2(e)\simeq e/2$ and $b_2(e) \simeq e/2$, then $\xi_2(e,\varpi) \simeq e/2$.
Hence, in this case, we get the eccentricity from the ratio of the low-order Fourier components;
\begin{eqnarray}
\label{eq:a2a1}
e \simeq {{2A_2}\over{A_1}}.
\end{eqnarray}
Similarly, it can be shown that
\begin{eqnarray}
\label{eq:a3a2}
e \simeq {{4A_3}\over{3A_2}}.
\end{eqnarray}

\begin{figure}
\begin{center}
\includegraphics[width=0.49\textwidth]{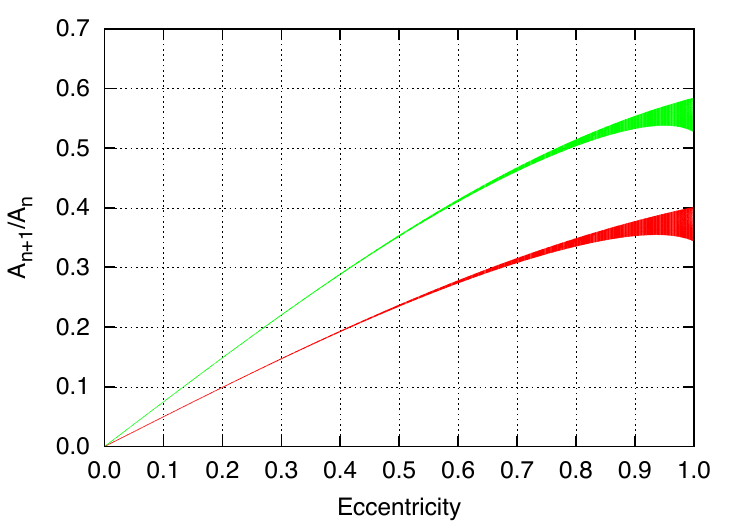}
\caption{Amplitude ratios of the harmonics, $A_{n+1}/A_n$, of time delays for $n=1$ (lower, red), and $n=2$ (upper, green), as functions of $e$. The values were numerically computed with Eq.\,\ref{eq:xi}. The band of each curve shows the range of $\varpi$ from 0 to $2\uppi$. The $\varpi$-dependence is very weak, hence we cannot determine $\varpi$ from the amplitude ratio. The angle $\varpi$ should be derived from Eq.\,\ref{eq:angle}, after obtaining a reasonable value of $e$ and the RV curve. The analytical ratios $A_2/A_1 \sim e/2$ and $A_3/A_2 \sim 3e/4$ for $e \ll 1$ are numerically validated.}
\label{fig:e_varpi}
\end{center}
\end{figure}

In more general cases, we need to solve equations \ref{eq:29} and \ref{eq:30}. The graphical solutions are shown in Fig.\,\ref{fig:e_varpi}.

\subsection{The eccentricity of KIC\,9651065}
\label{app:calculation}

The Fourier series applied to the time delays of KIC\,9651065 (\S\,\ref{ssec:rv}) gave amplitudes $A_1=165.57$, $A_2 = 38.41$ and $A_3 = 10.16$ ($\pm2.03$)\,s. Despite this star's moderately high eccentricity, it is instructive to see that the low eccentricity approximation still gives a sensible result. From Eqs\:\ref{eq:a2a1} and \ref{eq:a3a2} we calculate $e=0.464 \pm 0.026$ and $e=0.530 \pm 0.108$, respectively. The uncertainty-weighted mean of the two values is $e=0.468 \pm 0.025$.

Once $e$ and $\varpi$ are determined in a reasonably reliable way, such as with the low eccentricity approximation, all the coefficients $\xi_{n}(e,\varpi)$ and all the phases $\vartheta_{n}(e, \varpi)$ are uniquely determined with high precision. The RV curve is calculated with a series expansion of $\xi_{n}(e,\varpi)$. We found that $n=10$ is sufficiently high, with no noticeable difference between $n=10$ and $n=20$. Importantly, the high-order components are determined more precisely than is possible in practice through observational quantities.

As such, precise measurements of the coefficients of $A_1$, $A_2$, \mbox{$v_{\rm rad,1,max}$ [$= -c (d \tau/dt)_{\rm max}$]}, and \mbox{$v_{\rm rad,1,min} [= -c (d\tau/dt)_{\rm min}]$}, along with their uncertainties, are all that is required. The latter two quantities are later refined with the expansion of $\xi_{n}(e,\varpi)$. We tabulate the input values and the calculated orbital parameters for KIC\,9651065 in Table\:\ref{tab:9651065}.

\begin{table}
\begin{center}
\caption{Observed and calculated properties of the orbit of KIC\,9651065. Values of \mbox{$v_{\rm rad,1,max}$} and \mbox{$v_{\rm rad,1,min}$} are taken from the RV curve obtained by differentiating the time delays, and the uncertainties are approximated by the scatter.  $A_1$, $A_2$ and $\nu_{\rm orb}$ are found from the Fourier transform of the time delay data. The remaining values (below the mid-table rule) are calculated based on those observables. The mass function is calculated for $i=45^{\circ}$, and we have adopted $m_1 = 1.70$\,M$_{\odot}$ from \citet{huberetal2014}. $\phi_{\rm orb,periastron}$ is the orbital phase at periastron, and $T_{\rm periastron}$ is the corresponding time in BJD, at the epoch closest to the centre of the \kepler\ data set.}
\begin{tabular}{cccc}
\toprule
Quantity & Value & Units\\
 \midrule
$	v_{\rm rad,1,max,input}	$&$	20.2	\pm	2.5\phantom{0}	$ & km\,s$^{-1}$ \\
$	v_{\rm rad,1,min,input}	$&$	-13.5	\pm	3.2\phantom{-0} $& km\,s$^{-1}$\\
$	A_1	$&$	165.57	\pm	2.03\phantom{00}	$ & s\\
$	A_2	$&$	38.41	\pm	2.03\phantom{0}	$ & s\\
$	\nu_{\rm orb}	$&$	0.003667	\pm	0.000011	$ & d$^{-1}$\\
\midrule
$	P_{\rm orb}	$&$	272.70	\pm	0.82\phantom{00}	$ & d\\
$	e	$&$	0.468	\pm	0.025	$\\
$	\varpi	$&$	2.01	\pm	0.30	$ & rad\\
$	a_1 \sin i	$&$	0.37	\pm	0.02	$ & au\\
$	f(m_1,m_2,i=45^{\circ})	$&$	0.001667	\pm	0.000010	$ & M$_{\odot}$\\
$	m_1	$&$	1.70	\pm	0.17	$& M$_{\odot}$\\
$	m_2	$&$	0.26	\pm	0.02	$& M$_{\odot}$\\
$	\phi_{\rm orb,periastron} $&$ 0.46 \pm 0.03 $ & (0--1) \\
$ T_{\rm periastron} $&$ 2455400 \pm 15\phantom{00000} $ & d \\
\bottomrule
\end{tabular}
\label{tab:9651065}
\end{center}
\end{table}

\end{document}